\documentclass[aps,prl,twocolumn,superscriptaddress]{revtex4}

\usepackage{graphicx}
\usepackage{todonotes}
\newcommand{\be}{\begin{displaymath}}
\newcommand{\bn}{\begin{equation}}

\newcommand{\en}{\end{equation}}
\newcommand{\ee}{\end{displaymath}}

\usepackage{times}
\usepackage{amssymb}
\usepackage{amsmath}
\usepackage{url,hyperref}

\begin{document}
\title{Suppression of turbulence by trapped electrons in optimised stellarators}
%% OTHER IDEAS FOR TITLE
%Turbulence reduction through trapped electrons in stellarators 
%Possibility of enhanced confinement (through turbulence suppression) as seen in the Wendelstein 7-X stellarator
%
%
%Turbulence suppression in quasi-isodynamic stellarators
%
%
%
\author{J.~H.~E.~Proll}
\affiliation{Science and Technology of Nuclear Fusion, Department of Applied Physics, Eindhoven University of Technology, 5600 MB Eindhoven, The Netherlands}

\author{P.~Xanthopoulos}\author{P.~Helander}\author{G.~G.~Plunk}
\affiliation{Max-Planck-Institut f\"ur Plasmaphysik, Wendelsteinstra{\ss}e 1, 17491 Greifswald, Germany} 

\author{B.~J.~Faber}
\affiliation{HSX Plasma Lab, University of Wisconsin-Madison, Madison, Wisconsin 53706, USA} 
\affiliation{Department of Physics, University of Wisconsin-Madison, Madison, Wisconsin 53706, USA} 

\author{T.~G\"orler}
\affiliation{Max-Planck-Institut f\"ur Plasmaphysik, Boltzmannstra{\ss}e 2, 85748 Garching, Germany} 

\author{H.~M.~Smith}
\affiliation{Max-Planck-Institut f\"ur Plasmaphysik, Wendelsteinstra{\ss}e 1, 17491 Greifswald, Germany} \

\author{M.~J.~Pueschel}
\affiliation{Institute for Fusion Studies, The University of Texas at Austin, Austin, Texas 78712, USA}

\date{\today}

\begin{abstract}
In fusion devices, the geometry of the confining magnetic field has a significant impact on the instabilities that drive turbulent heat loss. This is especially true of stellarators, where the ?trapped electron mode? (TEM) is stabilised if specific optimisation criteria are satisfied, as in the Wendelstein 7-X experiment (W7-X). Here we find, by numerical simulation, that W7-X indeed has low TEM-driven transport, and also benefits from stabilisation of the ion-temperature-gradient mode, giving theoretical support for the existence of enhanced confinement regimes at finite density gradients.
\end{abstract}
\maketitle
\normalsize
\section{Introduction}
%can stay like this for now., done
In magnetic confinement fusion devices, there are usually three processes limiting the energy confinement: radiation losses, neoclassical transport which encompasses collisional diffusion---including the effect of particle drifts that arise due to gradients and curvature of the confining magnetic field---and turbulence. While tokamaks are never critically affected by collisional transport owing to their axisymmetry, stellarators historically suffered from poor confinement due to the lack of symmetry and the resulting high neoclassical transport losses. Optimised stellarators using the concepts of quasi-symmetry \cite{Nuhrenberg1988,Boozer}---like the quasi-helically symmetric experiment HSX \cite{Anderson1995}---or quasi-isodynamicity \cite{HN, Nuhrenberg2010}---like the recent superconducting stellarator Wendelstein 7-X \cite{Beidler1990,Klinger2013}---are designed to overcome the problem of large neoclassical transport \cite{Canik2007, Beidler2011}, rendering turbulence the dominant transport channel. 
As in tokamaks, ion-temperature-gradient modes (ITGs) and trapped-electron modes (TEM) have been identified as the most transport-relevant amongst the electrostatic modes. Recent research has focussed on studying the effects of the magnetic geometry available to stellarators on the instability of ITG and TEM.
For so-called quasi-isodynamic stellarators, one can argue that the second adiabatic invariant $J$ of fast-bouncing particles such as electrons is constant on flux surfaces with flux surface label $\psi$ and depends through the velocity $v$ on the total energy $E$ with $\partial J / \partial E >0$:
\begin{equation*}
J=\int m v_{\|}\mathrm{d}l = J(\psi, E)
\end{equation*}
If now an instability with frequency below the bounce frequency, $\omega\tau_b \ll 1$, moves a particle outwards by $\Delta \psi$,  the energy $\Delta E$ necessary for this follows from the conservation of $J$:
\begin{align*}
\Delta J = \frac{\partial J}{\partial \psi}\Delta \psi + \frac{\partial J}{\partial E}\Delta E = 0 \rightarrow \Delta E= -\frac{\partial J /  \partial \psi}{\partial J / \partial E}\Delta \psi
\end{align*}
%\begin{align*}
%\Delta J &= \frac{\partial J}{\partial \psi}\Delta \psi + \frac{\partial J}{\partial E}\Delta E = 0\\
%\Delta E&= -\frac{\partial J /  \partial \psi}{\partial J / \partial E}\Delta \psi
%\end{align*}
This means the movement is at the expense of the instability and thus is stabilising if $\partial J / \partial \psi <0$, i.e.~if $J$ has its maximum at the magnetic axis. 
In \cite{Proll2012, Helander2013} it was shown analytically that stellarators with the maximum-$J$-property should be stable to collisionless electron-driven TEM thanks to the stabilising property of the trapped electrons.
While the theory states that this resilience should only hold for perfectly quasi-isodynamic stellarators with the maximum-$J$ property, linear simulations \cite{Proll2013} showed that \mbox{W7-X}, which is only approximately quasi-isodynamic, indeed benefits from reduced TEM growth rates, too.
These results are however only linear and raise the question whether the enhanced stability actually results in less turbulent transport. Recent nonlinear analytical theory using the concept of available energy \cite{Helander2017} and preliminary simulations \cite{Helander2015} have hinted that this is indeed the case. 
The present paper investigates the question directly, demonstrating the fully nonlinear effect of trapped electron stabilisation on TEM turbulence, and also how the effect extends to ITG turbulence, by comparing simulation results obtained in \mbox{W7-X}, HSX and the \mbox{DIII-D} tokamak \cite{Luxon2002}.
\section{Simulation setup}
%--kicked out all grad-Te stuff, done\\
The collisionless electrostatic simulations were performed with the flux-tube version of the Eulerian code GENE \cite{Jenko2000, genewebsite}, which solves the gyrokinetic equation together with Maxwell's equations and incorporates realistic geometry when coupled to the GIST code \cite{Xanthopoulos2009}. 
For \mbox{W7-X} and HSX, the flux tubes studied cross the outboard midplane in the bean-shaped cross section of the stellarator at the midpoint of the flux tube), and in all three devices the flux surface chosen was at half normalized toroidal flux, i.e.~$s=\psi/\psi_0=0.5$. For more information about the geometry the reader is referred to \cite{Proll2013} for \mbox{DIII-D} and \mbox{W7-X} and to \cite{Proll2015} for HSX.  ITG turbulence was modeled with both adiabatic (ITG-ae) and kinetic (ITG-ke) electrons, and TEM with a pure density gradient; the resolution was chosen as seen in Table \ref{tab:resolution}. There, $nz$ refers to the number of grid points along a field line, $nkx$ the number of grid points in the radial direction, $nky$ to the number of Fourier modes in the binormal direction (i.e. perpendicular to both the field line and the radial direction), $nv$ to the number of grid points in the direction of parallel velocity $v_{\|}$,~$nw$ to that in the direction of the magnetic moment $\mu$, and $\mathrm{kymin}$ to the minimum value of the wave number in units of inverse gyroradius.
\begin{table*}[]
\centering
\begin{tabular}{c|c|c|c|c}
 &  ITG-ae&  ITG-ke&  $\nabla n$ TEM\\
 \mbox{DIII-D}&  192,64,64,48,20,0.05&  192,64,64,48,20,0.05&  192,48,64,48,20,0.05 \\
 HSX&  128,32,64,36,16,0.05&  384,72,64,36,8,0.05 & 192,48,64,36,16,0.01 \\
 \mbox{W7-X}& 384,64,128,48,20,0.05 & 256,64,96,48,10,0.05 & 256,64,96,48,10,0.05
\end{tabular}
\caption{Resolution for turbulence simulations (nkx, nky, nz, nv, nw, kymin) for ITGs with adiabatic (ITG-ae) and kinetic (ITG-ke) electrons and TEM with a pure density gradient.}
\label{tab:resolution}
\end{table*}
For TEM in HSX  $k_y \rho_s =0.1$ was found to be sufficient as a smallest binormal wave number. More details can be found in \cite{Faber2015}. The simulations are performed with a realistic mass ratio for hydrogen plasmas of $m_e/m_i=1/1836$ and a temperature ratio of $T_e/T_i=1$, where $m_{e}$ and  $m_{i}$ are the electron mass and the ion mass, respectively, and $T_{e}$ and $T_{i}$ the electron and ion temperatures, respectively.
\section{Nonlinear simulation results}
\subsection{Ion-temperature-gradient modes (ITG)}
\begin{figure}
\includegraphics[width=0.45 \textwidth]{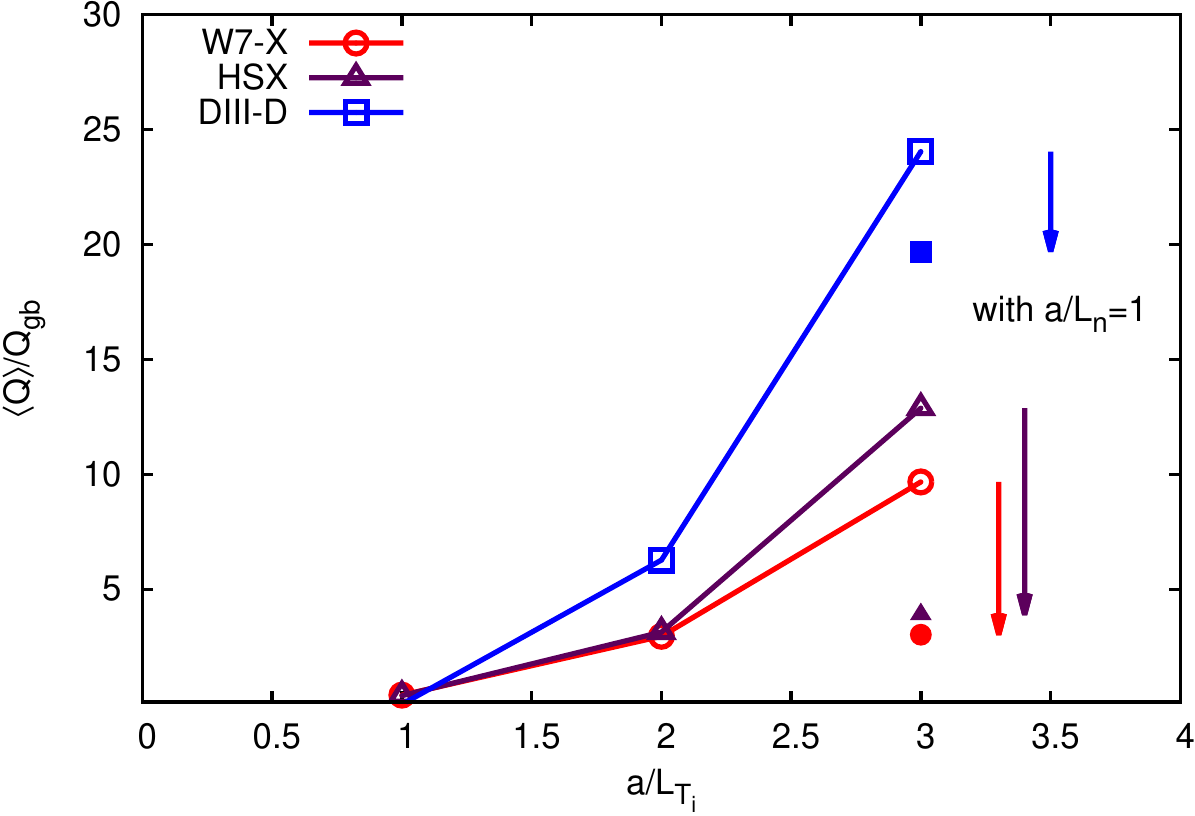}
\caption{\label{fig:ITGaQ} {\bf ITG adiabatic electrons:} Heat fluxes Q in units of $Q_{gb}=n_i T_i^{7/2}m_i^{-3/2}/(\Omega_i ^2 a^2)$ in \mbox{W7-X}, HSX and \mbox{DIII-D} for ion-temperature-gradient-driven turbulence with adiabatic electrons. The full symbols show how the heat fluxes change for an ion-temperature gradient of $a/L_{T_i}=3$ once a small density gradient $a/L_n=1$ is present.}
\end{figure}

\begin{figure}
\includegraphics[width=0.45 \textwidth]{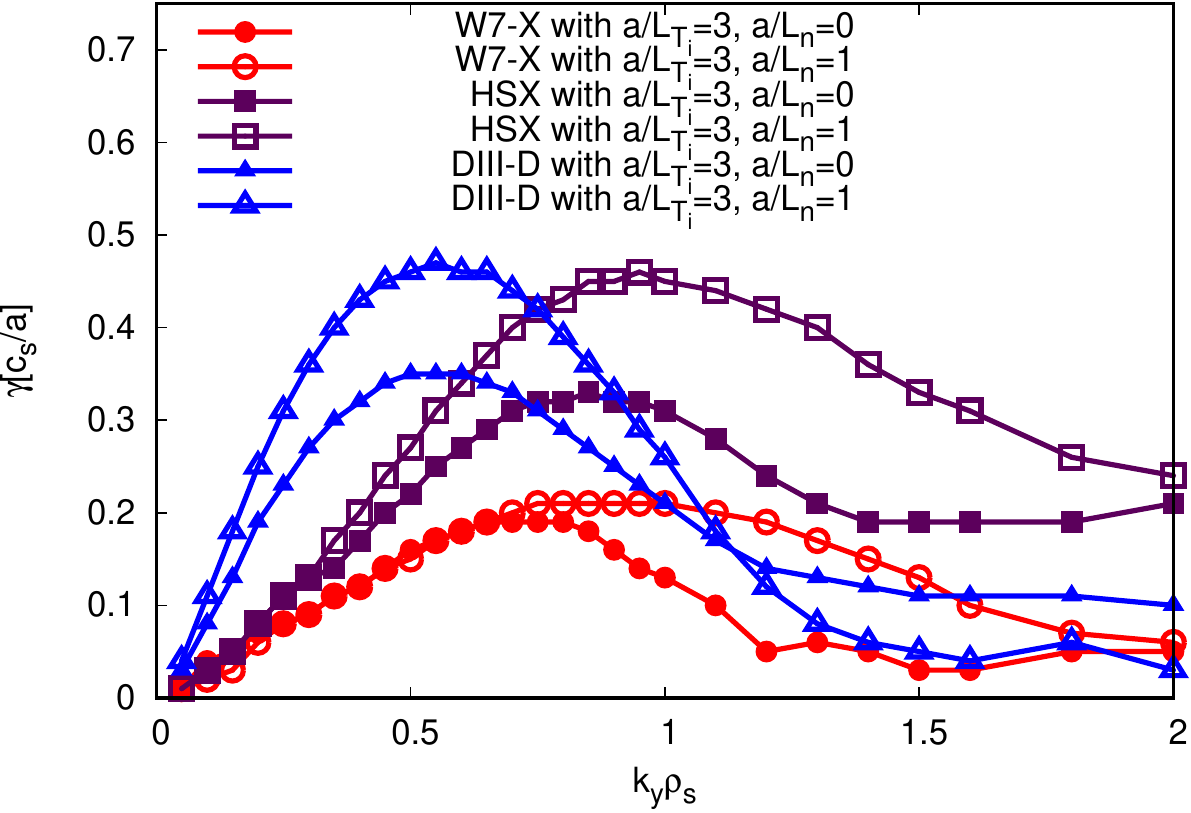}
\caption{\label{fig:ITGlinear}Growth rates $\gamma$ of ITGs with kinetic electrons (ITG-ke) at $a/L_{T_i}=3$ with (dashed lines) and without (solid lines) added density gradient of $a/L_n=1$  in \mbox{W7-X}, HSX and \mbox{DIII-D}.}
\end{figure}

First, we study the effect of the magnetic geometry on ITG turbulence by varying the ion temperature gradient $a/L_{T_i}$ while setting the electron-temperature gradient and the density gradient to zero. It is found that the heat flux (normalised to the gyro-Bohm value $Q_{GB}=n_i T_i^{7/2}m_i^{-3/2}/(\Omega_i ^2 a^2)$ with ion density $n_i$, ion temperature $T_i$, ion mass $m_i$, ion gyro frequency $\Omega_i$ and minor radius $a$) of turbulence resulting from ITGs with {\it adiabatic electrons} (ITG-ae) is smaller in both stellarators than in \mbox{DIII-D}, see Fig. \ref{fig:ITGaQ}. The difference can most likely be attributed to the generally smaller local curvature in \mbox{W7-X} and strong local shear and small global shear in both stellarators \cite{Plunk2014} and is also reflected in the linear growth rates (for \mbox{W7-X} and \mbox{DIII-D}, see \cite{Proll2013}). 
If we now add a density gradient $a/L_n=1$ for the case with $a/L_{T_i}=3$, we see the typical \cite{Coppi1967} stabilisation of ITG through a density gradient, resulting in reduced heat fluxes in all three devices. \\
Returning to the scenario with a flat density profile, the aforementioned inter-machine trend changes significantly if we also consider {\it kinetic electrons}: Initially we note that with kinetic electrons, all heat fluxes have increased, with the increase being strongest for \mbox{DIII-D} and weakest for HSX. This difference may be explained by the difference in trapped-particle fraction---\mbox{DIII-D} having the highest and HSX having the lowest trapped-particle fraction---as suggested by \cite{Redd1999}; the change in linear growth rates is also consistent with this explanation. \\
\begin{figure}
\includegraphics[width=0.45 \textwidth]{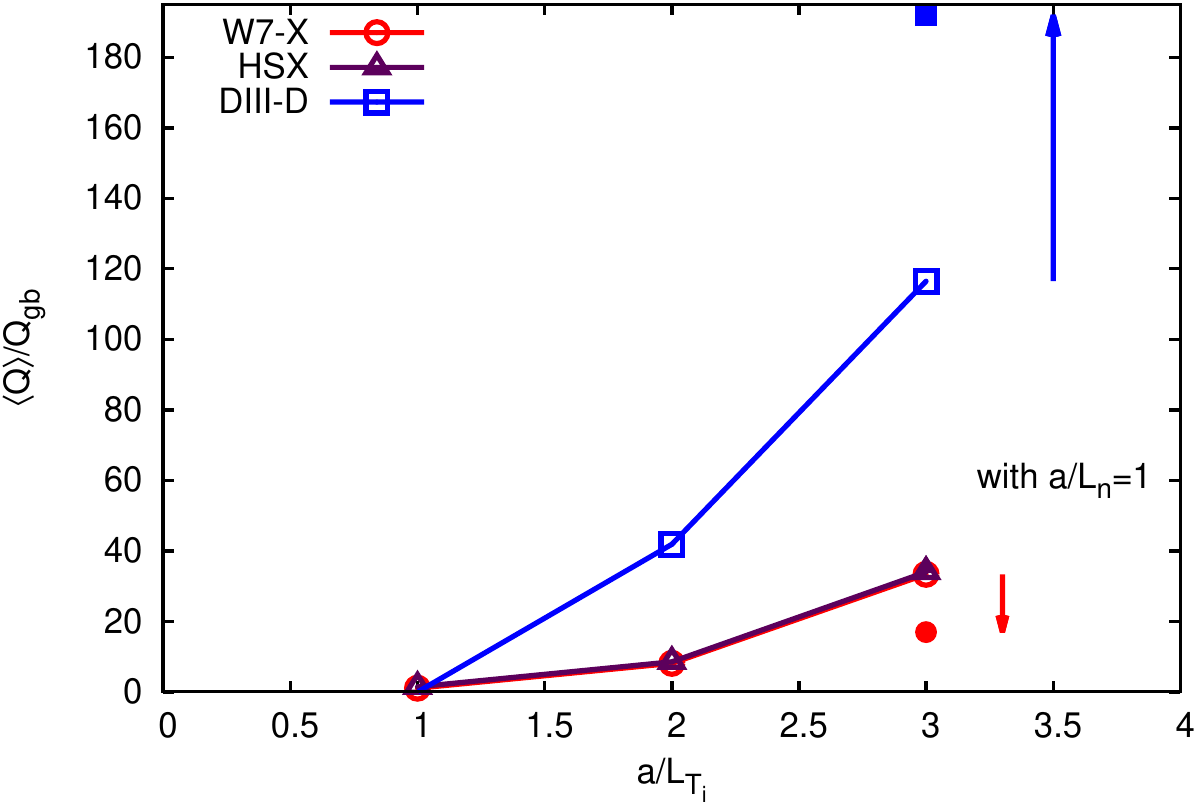}
\caption{\label{fig:ITGkQ}{\bf ITG kinetic electrons:} Heat fluxes Q in units of $Q_{gb}=n_i T_i^{7/2}m_i^{-3/2}/(\Omega_i ^2 a^2)$ in \mbox{W7-X}, HSX and \mbox{DIII-D} for ion-temperature-gradient-driven turbulence with kinetic electrons. The full symbols show how the heat fluxes change for an ion-temperature gradient of $a/L_{T_i}=3$ once a small density gradient $a/L_n=1$ is present. For HSX almost no change is observed.}
\end{figure}
Now if again a density gradient $a/L_n=1$ is added for the case with ion-temperature gradient $a/L_{T_i}=3$, we observe that the three devices behave very differently: 
For \mbox{DIII-D}, the heat flux increases strongly. This can be explained by trapped-electron-modes being destabilised (as is also seen in the linear growth rates, see Fig.\ref{fig:ITGlinear}), which then contribute to the heat flux. 
In \mbox{W7-X} we see the opposite behaviour: the heat flux decreases. This is rather remarkable, and to understand it one needs to consider the linear growth rates for the ITGs with and without a density gradient, see Fig.\ref{fig:ITGlinear}.
Only in \mbox{W7-X} do we see a reduction in linear growth rates once a density gradient is present, but not in \mbox{DIII-D} or in HSX. We attribute this to the maximum-$J$ property of \mbox{W7-X}, which leads to the electrons being stabilising, as was predicted analytically linearly \cite{Proll2012, Helander2013} and nonlinearly \cite{Helander2017}. Even though the theory focusses on TEM, one can argue that the enhanced stability extends to ITGs with a density gradient present. Indeed, it can be shown that less energy (in the electrons) is available to drive instabilities when the density profile is slightly peaked than when it is flat \cite{Helander2017}.
In HSX, no such linear stabilisation is observed but rather a destabilisation like in \mbox{DIII-D}---which is very well understandable as HSX is far from maximum-$J$---but the nonlinear heat flux does not increase accordingly. We thus note that in HSX another nonlinearly stabilising mechanism must be at play, such as enhanced energy transfer to stable eigenmodes \cite{Terry2018, Hegna2018}.  
\subsection{Trapped-electron modes (TEM)} 
\begin{figure}
\includegraphics[width=0.45 \textwidth]{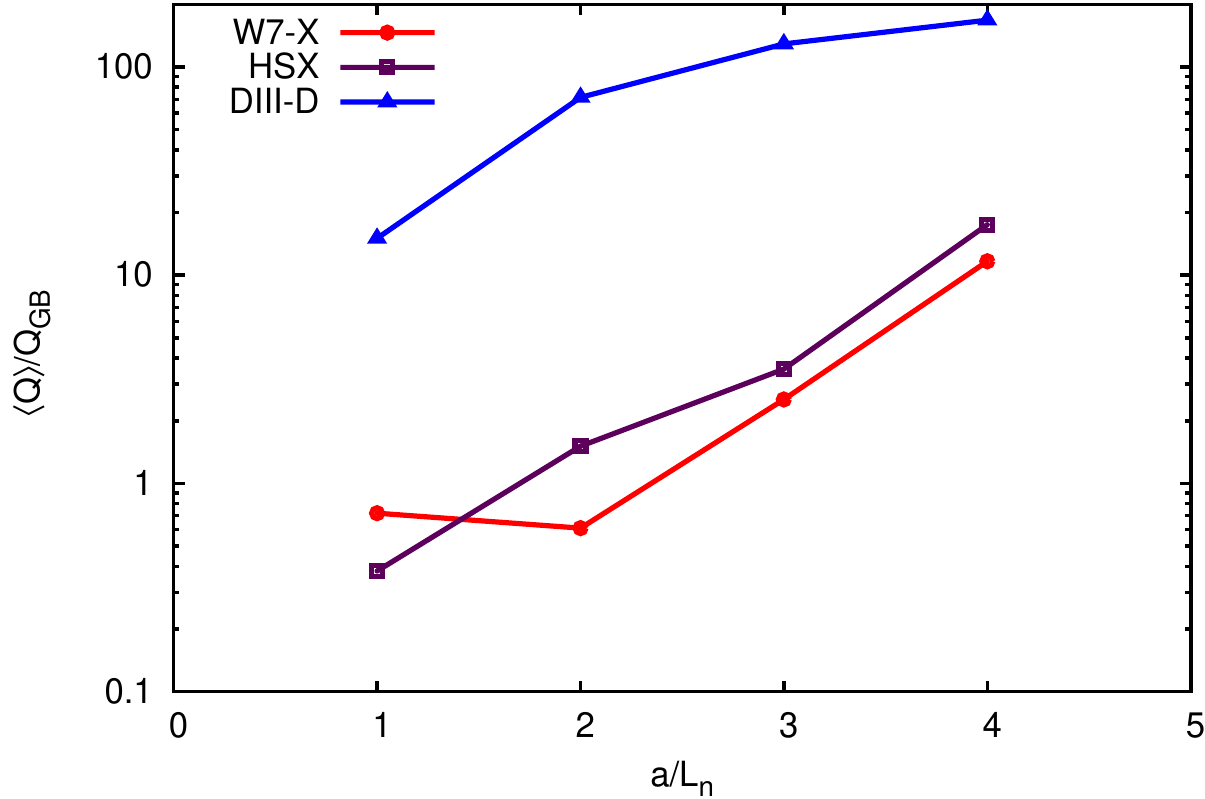}
\caption{\label{fig:TEMQ}{\bf TEM:} Heat fluxes Q in units of $Q_{gb}=n_i T_i^{7/2}m_i^{-3/2}/(\Omega_i ^2 a^2)$ in \mbox{W7-X}, HSX and \mbox{DIII-D} for density-gradient-driven turbulence with kinetic electrons.}
\end{figure}

For the TEM studies both temperature gradients are set to zero and only the density gradient is varied. The heat fluxes of both stellarators are up to two orders of magnitude smaller than that in \mbox{DIII-D} (see Fig. \ref{fig:TEMQ}). 
In \mbox{W7-X}, the explanation for the low heat flux lies again with the maximum-$J$ property: not only are the linear growth rates of the density-gradient-driven TEM much smaller than in the two other devices (not shown here)---as predicted in \cite{Proll2012, Helander2013}---but we conclude that the enhanced stability also holds nonlinearly. 
One surprising observation is that in \mbox{W7-X} the heat flux at $a/L_n=1$ is slightly larger than that at $a/L_n=2$ (Fig. \ref{fig:TEMQ}). This cannot be explained by the difference in linear growth rates (not shown here), which increase monotonically with increasing density gradient, as expected. In these linear simulations, we do however observe a distinctly different mode at large scales, with a real frequency in the ion-diamagnetic direction, whereas for both $a/L_n=2$ and $a/L_n=3$ the modes at these large scales were propagating in the electron diamagnetic direction. This different mode---possibly the the ion-driven trapped electron mode (iTEM) recently described by Plunk {\it et al.} in \cite{Plunk2017}---may have different nonlinear properties and therefore lead to the unexpectedly high heat flux.\\
For HSX, the low heat flux is not as readily explained. Just as for the ITG simulations with a density gradient present, the growth rates are much higher than in \mbox{W7-X} due to the lack of maximum-$J$-property. Even though the TEM in HSX are of smaller scales than in \mbox{DIII-D}, a simple quasi-linear theory or mixing-length estimate cannot explain why the heat flux is almost as low in HSX as in \mbox{W7-X}.
Again, we suspect that a different turbulence saturation mechanism is at work, e.g. through coupling to damped eigenmodes.
Note that similar nonlinearly stabilising features have been seen in ITG turbulence comparisons between HSX and an axisymmetric configuration \cite{McKinney2019}.\\
%\textcolor{red}{comment on disparity beween linear and NL ITG/TEM?\\
%Another surprising observation in \mbox{W7-X} is the large disparity between the heat fluxes for ITGs and TEMs in spite of the comparable linear growth rates. This is also the case for HSX. There, however, it can be explained by the fact that the most unstable modes for the ITG case are at larger scales than for the TEM case, which could lead to a higher ITG heat flux. For \mbox{W7-X} on the other hand, the scales for linear TEMs and ITGs are very similar, and the reason for the different heat fluxes is less clear. 
%A more detailed analysis of these findings will be presented in a longer paper.\\ }
\begin{figure}
\includegraphics[width=0.45 \textwidth]{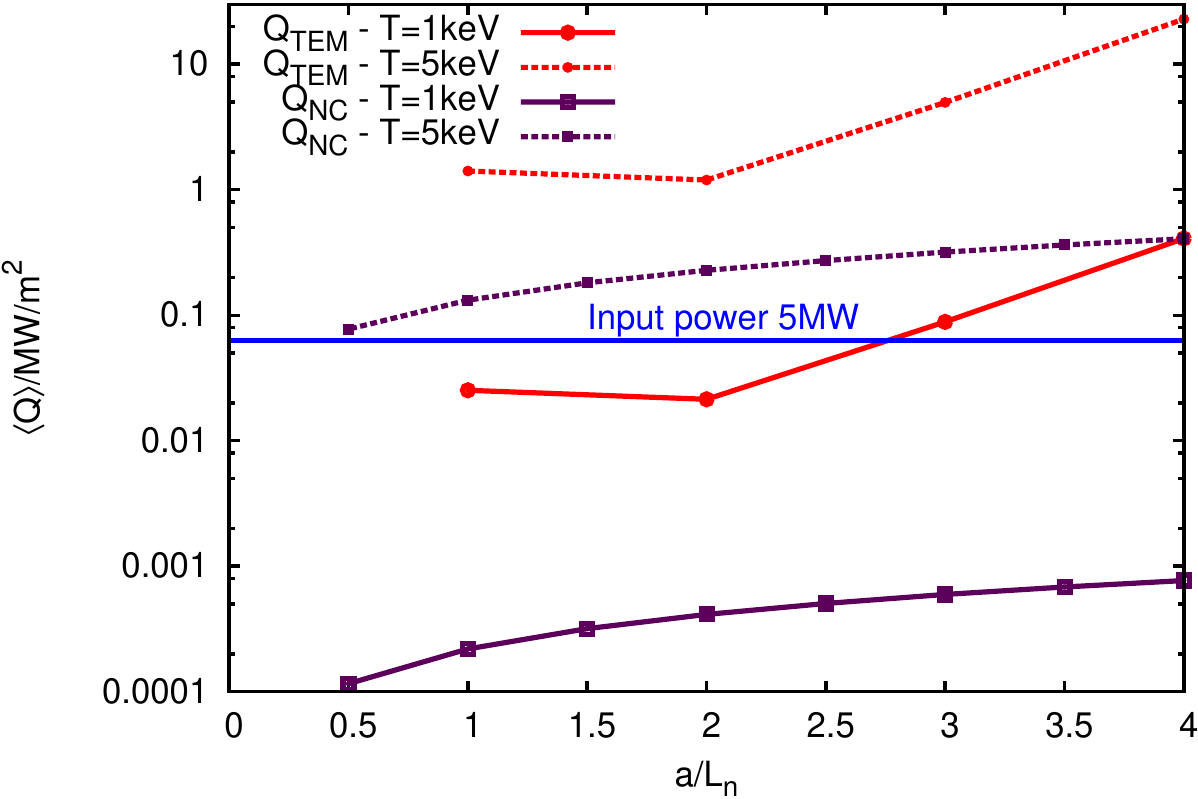}
\caption{\label{fig:TEMNC}TEM turbulent heat flux $Q_{TEM}$ and neoclassical heat flux $Q_{NC}$, respectively, in \mbox{W7-X} for different density gradients. The usual normalisation in gyro-bohm units has been transformed into $MW/m^2$, assuming a density of $n=5 \cdot 10^{19}/m^3$ and temperatures of $T = 1\mathrm{keV}$ and $T = 5\mathrm{keV}$. For comparison, the typical input heating power for \mbox{W7-X} of about $5 MW/m^2$ is given.}
\end{figure}
One final observation regarding \mbox{W7-X} is that, in spite of the TEM heat flux being small compared with the ITG heat flux or the heat fluxes in \mbox{DIII-D}, it is still large compared with the neoclassical flux as calculated by the SFINCS code \cite{Landreman2014} at typical values of density $n=10^{20}/m^3$ and temperature $T=1\mathrm{keV}-5\mathrm{keV}$ (see Fig. \ref{fig:TEMNC}). 
This clearly supports turbulent transport as the dominant transport channel in \mbox{W7-X}, and is very much in line with the experimental observations \cite{Dinklage2018}, where it was found that the heat flux calculated by neoclassical theory is not sufficient to explain experimental measurements. \\
In summary, we find that owing to the maximum-$J$ property of \mbox{W7-X}, the turbulent heat flux of both density-gradient-driven TEM and ITGs with a small density gradient is much lower than in a tokamak, while HSX seems to benefit from a more powerful saturation mechanism, despite lacking maximum-$J$ optimisation. This suggests that turbulence in stellarators like \mbox{W7-X} or HSX can be suppressed by increasing the density gradient for example through pellet injection, as recently reported for \mbox{W7-X} \cite{Bozhenkov2019}, illuminating a path toward enhanced fusion performance.
\begin{acknowledgments}
The authors would like to thank the GENE team - most notably D. Told, F. Jenko, H. Doerk and A. Ba\~n\'on-Navarro - for their efforts and support. We would also like to thank G. Hammett, S. Lazerson, H. Mynick and J. Citrin for lots of stimulating and helpful discussions. The simulations were performed on MPCDFs Hydra and Helios at IFERC-CSC. 
This work has been carried out within the framework of the EUROfusion Consortium and has received funding from the Euratom research and training programme 2014-2018 and 2019-2020 under grant agreement No 633053. The views and opinions expressed herein do not necessarily reflect those of the European Commission.
\end{acknowledgments}

\end{document}